\documentclass{aastex}
\usepackage{emulateapj5}
\usepackage{apjfonts}
\usepackage{epsf}
\usepackage{onecolfloat}

\newcommand{\beq}{\begin{equation}}
\newcommand{\eeq}{\end{equation}}
\newcommand{\beqa}{\begin{eqnarray}}
\newcommand{\eeqa}{\end{eqnarray}}



\begin{document}
\twocolumn[
\vspace{-10mm}
\title{On the Origin of the Global Schmidt Law of Star Formation}

\author{Andrey V. Kravtsov}

\affil{Department of Astronomy and
Astrophysics and Center for Cosmological Physics,\\
5640 S. Ellis Ave., The University of Chicago, Chicago IL 60637\\
e-mail: {\tt andrey@oddjob.uchicago.edu}}


\begin{abstract}
  
  One of the most puzzling properties of observed galaxies is the
  universality of the empirical correlation between the star formation
  rate and average gas surface density on kiloparsec scales (the
  Schmidt law). In this study I present results of self-consistent
  cosmological simulations of high-redshift galaxy formation that
  reproduce the Schmidt law naturally, without assuming it, and
  provide some clues to this puzzle. The simulations incorporate the
  main physical processes critical to various aspects of galaxy
  formation and have a dynamic range high enough to identify
  individual star forming regions. The results indicate that the
  global Schmidt law is a manifestation of the overall density
  distribution of the interstellar medium (ISM). In particular, the
  density probability distribution function (PDF) in the simulated
  disks is similar to that observed in recent state-of-the-art
  modeling of the turbulent ISM and has a well-defined generic shape.
  The shape of the PDF in a given region of the disk depends on the
  local average surface density $\Sigma_{\rm g}$.  The dependence is
  such that the fraction of gas mass in the high-density tail of the
  distribution scales as $\Sigma_g^{n-1}$ with $n\approx 1.4$, which
  gives rise to the Schmidt-like correlation. The high-density tail of
  the PDF is remarkably insensitive to the inclusion of feedback and
  details of the cooling and heating processes.  This indicates that
  the global star formation rate is determined by the supersonic
  turbulence driven by gravitational instabilities on large scales,
  rather than stellar feedback or thermal instabilities on small
  scales.

\end{abstract}


]
\section{Introduction}
\label{sec:intro}

Formation of galaxies is a complicated process involving a variety of
physical phenomena on a large range of spatial and temporal scales.
Although hierarchical structure formation scenario proved to be a very
successful framework for interpreting a wide variety of observational
data, much remains to be understood about the processes that shape
properties of galaxies.  In particular, star formation is one of the
most important, yet very poorly understood processes.

On small scales ($\lesssim 10-100$~pc), star formation appears to be a
complicated and stochastic process.  Understanding how stars and
stellar clusters form in individual molecular clouds is still
sketchy, even in the best studied cases \citep[e.g.,][]{hartmann02}.
It is therefore remarkable that on kiloparsec scales star formation
exhibits surprising regularity.  \citet{schmidt59} first argued that
the star formation rate (SFR) in galaxies is a power law function of
average gas density. This simple `Schmidt' law
\begin{equation}
\Sigma_{\rm SFR}\propto \Sigma^{n}_{\rm g}
,
\label{eq:schmidtlaw}
\end{equation} 
where $\Sigma_{\rm SFR}$ and $\Sigma_{\rm g}$ are the surface
densities of young stars and gas averaged on a large ($\sim$~kpc)
scale, was found to hold for a wide range of galaxy types, star
formation rates, and gas densities
\citep{kennicutt83,kennicutt89,kennicutt98araa,kennicutt98}.  The star
formation histories of the Local Group dwarfs indicate that the
Schmidt law applied at high redshifts \citep{gnedin00}.  The power-law
index of the correlation falls in the range $n\approx 1.3-1.6$,
depending on the tracers used and the scales considered
\citep{kennicutt98araa}. In addition, star formation rate in galaxies
declines sharply below a critical threshold surface density of $\sim
5-10\ \rm M_{\odot}\,pc^{-2}$ \citep[e.g.,][and references
therein]{kennicutt89,martin_kennicutt01}.  This density is thought to
be associated with large-scale gravitational and shearing stability
thresholds, although details of the mechanism are still uncertain.
Nevertheless, at high gas densities the form of the Schmidt law
appears to be remarkably consistent from galaxy to galaxy, both in
terms of slope and absolute efficiency (i.e., normalization of the
relation).

Despite the apparent simplicity, a unique explanation for the Schmidt
law proved to be elusive. To a large extent, the problem is that the Schmidt 
law can be explained by any process in which gas consumption time
depends on the local average dynamical time ($\propto \rho^{-1/2}$). 
A wide variety of possible models with such scaling have been 
proposed in the last three decades \citep[see][ for a review]{elmegreen02}.
Nevertheless, the universality of the star formation efficiency and the slope
of the Schmidt law in the face of complexity of star formation on small scales
remains a puzzle. In this {\it Letter} I present results of self-consistent
cosmological simulations of high-redshift galaxy formation that
reproduce the Schmidt law naturally, without assuming it, and provide some
clues to this puzzle. 

\section{Implementation of star formation}
\label{sec:sf}

The simulations presented in this paper were performed using the
Eulerian gasdynamics$+N$-body Adaptive Refinement Tree (ART) code.
The code is based on the cell-based approach to adaptive mesh
refinement (AMR) developed by \citet{khokhlov98}.  The algorithm uses
a combination of multi-level particle-mesh
\citep{kravtsov_etal97,kravtsov99} and shock-capturing Eulerian
methods \citep{vanleer79,colella_glaz85} to follow the evolution of
the DM and gas, respectively. High dynamic range is achieved by
applying adaptive mesh refinement to both gasdynamics and gravity
calculations.

Several physical processes critical to various aspects of galaxy
formation are implemented in the code: star formation,
metal enrichment and thermal feedback due to the supernovae type II
and type Ia (SNII/Ia), self-consistent advection of metals,
metallicity- and density-dependent cooling and UV heating due to
cosmological ionizing background using cooling and heating rates
tabulated for the temperature range $10^2<T<10^9$~K and a grid of
densities, metallicities, and UV intensities using the {\tt Cloudy} code
\citep[ver. 96b4,][]{ferland_etal98}. The cooling and heating rates take into
account Compton heating/cooling of plasma, UV heating, atomic and
molecular cooling.  The detailed implementation of these processes
will be described elsewhere \citep{kravtsov03b}. As I will show below,
the results presented in this paper are not sensitive to the details of these
processes.  Here I will focus on the implementation of star formation,
the process which is the subject of this study.

In numerical simulations star formation is usually assumed to occur
in certain ``star forming'' regions identified using some specific criteria.
The gas in such regions is then converted  into stars 
on a characteristic gas consumption time scale, $\tau_{\ast}$:
$\dot{\rho}_{\ast}=\rho_{\rm g}/\tau_{\ast}$.
Observationally, the efficiency of star formation on small ($\lesssim
100$~pc) scales in the densest regions of the ISM is significantly
higher than the low global efficiency indicated by the Schmidt law
\citep[e.g.,][]{elmegreen02}.  This means that the value of
$\tau_{\ast}$ and, perhaps more importantly, its density dependence in
simulations should be different depending on whether star forming
regions are resolved. Clearly, the use of the Schmidt law is
observationally motivated only on scales $\gtrsim 1$~kpc. Although
implementations vary, galaxy formation simulations usually adopt gas
consumption time that scales as $\tau_{\ast}\propto\max(t_{\rm
  cool},t_{\rm dyn})$, where $t_{\rm cool}$ and $t_{\rm dyn}$ are the
local cooling and dynamical time, respectively.  In even moderately
dense regions, $t_{\rm cool}\ll t_{\rm dyn}$ and $\tau_{\ast}\propto
t_{\rm dyn}\propto \rho^{-1/2}$, which results in the Schmidt-like
star formation law: $\dot{\rho}_{\ast}\propto\rho_{\rm g}^{1.5}$. 

In this study, a ``minimal'' star formation prescription was used.
Namely, the stars were formed with a {\em constant} $\tau_{\ast}$ so
that $\dot{\rho}_{\ast}\propto \rho_{\rm g}$. This ``constant
efficiency'' model on the scale of star forming regions is
well motivated by observations
\citep[e.g.,][]{young_etal96,wong_blitz02}.  The star formation was
allowed to take place only in the coldest and densest regions,
$T<T_{\rm SF}$ and $\rho_{\rm g} > \rho_{\rm SF}$, but no other
criteria (like the collapse condition $\nabla\cdot {\bf v} < 0$) were
imposed. I used $\tau_{\ast}=4$~Gyrs, $T_{\rm SF}=9000$~K, and
$\rho_{\rm SF}=1.64{\ \rm M_{\odot}pc^{-3}}$ or atomic hydrogen number
density of $n_{\rm H}=50{\ \rm cm^{-3}}$.  The adopted values of
$T_{\rm SF}$ and $\rho_{\rm SF}$ are quite different from the typical
temperatures and density of star forming molecular cores: $T\lesssim
30-50$~K and $n_{\rm H}\gtrsim 10^4{\ \rm cm^{-3}}$.  They are,
however, more appropriate to identify star forming regions on $\sim
100$~pc scales which are resolved in the presented simulations. In
practice, $T_{\rm SF}$ is not relevant because most of the gas at
$\rho>\rho_{\rm SF}$ is at temperatures of just a few hundred degrees
Kelvin.  The adopted gas consumption time scale $\tau_{\ast}$ is
derived from the observed normalization of the Schmidt law.

Algorithmically, star formation events are assumed to occur once every
global time step $\Delta t_0\lesssim 10^7$ yrs, the value close to the
observed time scales \citep[e.g.,][]{hartmann02}.  
Collisionless stellar particles with mass
$m_{\ast}=\dot{\rho}_{\ast}\Delta t_0$ are formed in every unsplit
mesh cell during star formation events. The mass of stellar particles
is restricted to be larger than $10^4{\ \rm M_{\odot}}$ but
smaller than 2/3 of the gas mass contained in the parent cell. This is
done in order to keep the number of stellar particles computationally
tractable while avoiding a sudden dramatic decrease in the local gas
density.
  
Once formed, each stellar particle is treated as a single-age stellar
population and its feedback on the surrounding gas is implemented
accordingly. I use the word feedback in a broad sense to include
injection of energy and heavy elements (metals) via stellar winds and
supernovae and secular mass loss.  Specifically, in the simulations
analyzed here, I assumed that the stellar initial mass function (IMF) is
described by the \citet{miller_scalo79} functional form with stellar
masses in the range $0.1-100\ \rm M_{\odot}$. All stars with 
$M_{\ast}>8{\ \rm M_{\odot}}$ deposit $2\times 10^{51}$~ergs of
thermal energy and a mass $f_{\rm Z}M_{\ast}$ 
of heavy elements  in their parent cell (no delay of cooling was
  introduced in these cells). The metal fraction is $f_{\rm Z}= {\rm
  min}(0.2,0.01M_{\ast}-0.06)$, which crudely
approximates the results of \citet{woosley_weaver95}. In addition, the
stellar particles return a fraction of their mass and metals to the
surrounding gas at a secular rate $\dot{m}_{\rm
  loss}=m_{\ast}\,\,C_0(t-t_{\rm birth} + T_0)^{-1}$ with $C_0=0.05$ and
$T_0=5$~Myr \citep{jungwiert_etal01}. The code also accounts for SNIa
feedback assuming a rate that slowly increases with time and broadly
peaks at the population age of 1~Gyr. I assume that a fraction of
$5\times 10^{-3}$ of mass in stars between 3 and $8\ \rm M_{\odot}$
explodes as SNIa over the entire population history and each SNIa
dumps $2\times 10^{51}$ ergs of thermal energy and ejects $1.3\ \rm
M_{\odot}$ of metals into parent cell. For the assumed IMF, 75 SNII
(instantly) and 11 SNIa (over several billion years) are produced by a
$10^4\ \rm M_{\odot}$ stellar particle. 

\section{Numerical Simulations}
\label{sec:sim}

In this paper I present two simulations of the early ($z\geq 4$) stages
of evolution for a galaxy of typical mass: $\approx 10^{12}h^{-1}\ \rm
M_{\odot}$ at $z=0$. At the analyzed epochs, the galaxy has already built
up a significant portion of its final mass: $1.3\times 10^{10}h^{-1}\ 
\rm M_{\odot}$ at $z=9$ and $2\times 10^{11}h^{-1}\ \rm M_{\odot}$ at
$z=4$. The evolution is started from a random realization of a
gaussian density field at $z=50$ in a periodic box of $6h^{-1}\ \rm Mpc$
with an appropriate power spectrum
and is followed assuming flat $\Lambda$CDM model:
$\Omega_0=1-\Omega_{\Lambda}=0.3$, $\Omega_b=0.043$, $h=H_0/100=0.7$,
$n_s=1$, and $\sigma_8=0.9$. The parameters have their usual meaning
and are consistent with recent cosmological constraints.

To increase the mass resolution, a low-resolution simulation was run
first and a galactic-mass halo was selected. A lagrangian region
corresponding to five virial radii of the object at $z=0$ was then
identified at $z=50$ and re-sampled with additional small-scale
waves \citep{klypin_etal01}.  The total number of DM particles in the
high-resolution lagrangian region is $2.64\times 10^6$ and their mass
is $m_{\rm DM}=9.18\times 10^5h^{-1}\ \rm M_{\odot}$. Outside the
high-resolution region the matter distribution was sampled with
$\approx 3\times 10^5$ higher mass particles.

As the matter distribution evolves, the code adaptively and
recursively refines the mesh in high-density regions. In the simulations I
present, two main refinement criteria were used: gas and DM mass in
each cell.  The code used a uniform $64^3$ grid to cover the entire
computational box. The lagrangian region, however, was always
unconditionally refined to the third refinement level, corresponding
to the effective grid size of $512^3$. Beyond the third level, a mesh
cell was tagged for refinement if its gas {\em or} DM mass exceeded
0.125 and 0.0625 times the mean mass expected for the average density
in each component in the zeroth level (i.e., uniform grid) cell,
respectively. The refinement thus follows the collapse of $1.2\times
10^6h^{-1}\ \rm M_{\odot}$ (gas) and $3.7\times 10^6h^{-1}\ \rm
M_{\odot}$ (DM) mass elements in a quasi-lagrangian fashion. 

The maximum allowed refinement level $l_{\rm max}$ was set to nine.  A
total of $\approx 1.1\times 10^7$ mesh cells was used at $z=4$ with
$\approx 2.5\times 10^5$ of them at refinement levels of 8 and 9.  The
volume of high-density cold star forming disks forming in DM halos was
refined to $l_{\rm max}=9$. The physical size of mesh cells in the
simulations was $\Delta x_l=26.16\,\,[10/(1+z)] 2^{9-l}\ \rm pc$, where
$l$ is the cell's level of refinement.  Each refinement level was
integrated with its own time step $\Delta t_l=\Delta t_02^{-l}\approx
2\times 10^4\,\, 2^{9-l}\ \rm yrs$, where $\Delta t_0\lesssim 10^7\ 
\rm yrs$ is the global time step on the zeroth level, 
set using the CFL condition.

The two analyzed simulations are the same in all respects, except that
one included all of the feedback processes described above (energy and
metal injection, mass loss, etc.), while the other did not. The
simulation with no feedback is designated as NF throughout this paper.
Note that the cooling rates in the run with feedback accounted
for the local metallicity of the gas, while in the run with no
feedback the significantly lower zero-metallicity cooling rates were
used.

\begin{figure}[t]
\centerline{\epsfxsize=3.5truein\epsffile{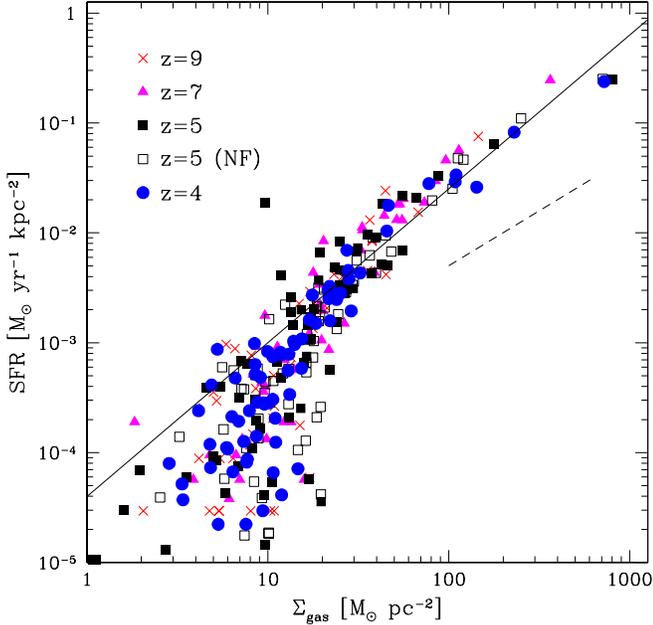} }
\caption{\footnotesize Star formation rate surface density vs. gas surface
  density in the simulated galaxies at different epochs. The open
  squares show $z=5$ epoch in the simulation in which no feedback (NF)
  processes were included.  The rates and gas density are averaged
  over $3.35\,\,[(1+z)/10]$~kpc. The solid line is relation ${\rm
    SFR}\propto \Sigma_{\rm g}^{1.4}$ while the dashed line is ${\rm
    SFR}\propto \Sigma_{\rm g}$. }
\label{fig:simschmidt}
\vspace{-0.5cm}
\end{figure}

\section{Results}
\label{sec:results}

Figure~\ref{fig:simschmidt} shows the star formation rate per
area vs.  gas surface density at different epochs in the simulations.
Individual points correspond to the level $l=2$ cells of size
$\Delta x_2=3.35\,\,[(1+z)/10]$~kpc. Note that the cells represent not
a single object but several 
dozen galaxies with masses $\gtrsim 10^8h^{-1}\ \rm M_{\odot}$ distributed
in a high resolution $\sim 1$~Mpc (comoving) lagrangian region. 
The gas surface density was
computed simply as $\Sigma_{\rm gas}=\rho_{\rm g}\Delta x_2$ and the
SFR in each cell was estimated as $m_{\ast}t_{\ast}^{-1}\Delta
x_2^{-2}$, where $m_{\ast}$ is the mass of stars younger than
$t_{\ast}=3\times 10^7$~yrs.  The averaging scale and $t_{\ast}$ are
close to the values typical for observational estimates
\citep[see][]{kennicutt98araa}. The resulting SFR per area and
$\Sigma_{\rm gas}$ are not sensitive to variation in averaging
scale and $t_{\ast}$ over a large range of values. 

The figure shows that the star formation rate is strongly correlated
with gas surface density: ${\rm SFR}\propto \Sigma_{\rm gas}^n$ with
$n\approx 1.4$ (solid line) for $\Sigma_{\rm gas}\gtrsim 10\ \rm
M_{\odot}pc^{-2}$.  At smaller densities the rate rapidly decreases
and most $l=2$ cells with $\Sigma_{\rm gas}\lesssim 5\ \rm
M_{\odot}pc^{-2}$ do not contain young stars (i.e., ${\rm SFR}=0$).
The simulations thus reproduce the slope of the empirical Schmidt law
and the observed critical threshold for star formation remarkably
well.  Figure~\ref{fig:simschmidt} also shows that the relation is
universal: it does not depend sensitively on redshift or input
physics. The correlation is not sensitive to the variations of
averaging scale and $t_{\ast}$ by at least a factor of four. This is a
non-trivial result because the correlation is observed on kiloparsec
scales, while star formation occurs in individual $20-50$~pc cells
with the {\em rate simply proportional to the gas density}:
$\dot{\rho}_{\ast}\propto \rho_{\rm g}$. The ${\rm SFR}\propto
\Sigma_{\rm g}$ correlation is indeed recovered when averaging is done
on the scale of individual star forming regions ($\sim 100$~pc). The
observed scaling ${\rm SFR}\propto \Sigma_{\rm gas}^{1.4}$ on kpc
scales is therefore reproduced naturally when a constant efficiency of
star formation is assumed on the scale of molecular complexes.

The scatter about the mean of the correlation is rather small ($\sim
0.2-0.3$~dex) and is well below the observed scatter. The observed
individual rates and densities, however, have errors $\gtrsim
0.2-0.3$~dex so larger scatter can be expected.  The scatter at any
given epoch is even smaller, which is remarkable given that the
galaxies at $z\gtrsim 4$ undergo very frequent and violent mergers.
The normalization of the correlation is somewhat lower than
observed. This can be fixed by lowering
$\tau_{\ast}$. The gas consumption time is a free
parameter of the model and so is the normalization of the ${\rm
  SFR}-\Sigma_{\rm g}$ correlation.

\section{Discussion and Conclusions}
\label{sec:conclusions}

\begin{figure}[t]
\centerline{\epsfxsize=3.5truein\epsffile{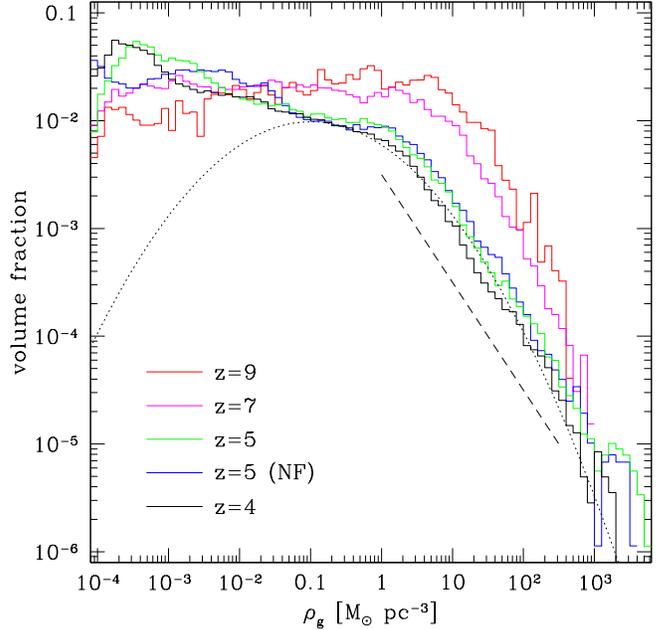} }
\caption{\footnotesize Probability distribution function of gas density
  in the simulated gas disks at different redshifts. The PDF,
  the fraction of total volume in a given density interval
  ($\Delta\log_{10}\rho_{\rm g}=0.1$), was computed in the
  regions refined to $l\geq 8$ refinement level. The PDFs are normalized
to unity. The histograms show 
 outputs at different epochs; the blue histogram shows results for $z=5$
output of the run without feedback (NF). The dashed line is $\propto \rho_{\rm g}^{-1}$, while the dotted line shows a log-normal PDF }
\label{fig:dpdfz}
\vspace{-0.5cm}
\end{figure}

What mechanism gives rise to the Schmidt law in simulated galaxies and
what explains its universality?  Some understanding can be gained by
considering the density distribution of gas in simulated galaxy disks.
Figure~\ref{fig:dpdfz} shows the probability distribution function
(PDF) of gas density in the disks at different epochs. Although the
PDF evolves with time, its general shape at every epoch can be
characterized by the relatively flat region at $\rho_{\rm g}\lesssim
1-10\ \rm M_{\odot}pc^{-3}$ and a power law distribution at high
densities.  The high-density tail is also well approximated by the
log-normal distribution (dotted line in Fig.~\ref{fig:dpdfz}). The
evolution is driven primarily by shocks associated with supersonic
turbulence induced by gravitational and shearing instabilities and,
occasionally, by tidal interactions during mergers in the gas disks.
For instance, the most massive disk in the simulation at $z=4$ has
well-developed large-scale spiral arms and exhibits signs of two minor
mergers in progress \citep{kravtsov03b}.

Stellar feedback appears to have only a minor effect on the overall
shape of the PDF.  Comparison of PDFs for the feedback and no-feedback
(NF) runs at $z=5$ shows that distribution is remarkably insensitive
to the inclusion of several feedback processes. The only difference is
at the lowest gas densities, where the simulation with feedback has
considerably more low-density hot gas due to the energy injection by
SNe. Note that in terms of morphology of gas distribution the effect
of energy injection is significant as the lowest density regions
occupy a large fraction of the volume.  Globally, the effect of energy
feedback on the thermal state of gas in the halos and surrounding
intergalactic medium is also significant \citep{tassis_etal03}. This
makes the insensitivity of the high-density PDF to feedback even more
remarkable. The distributions are also not sensitive to the
differences in metallicity and cooling rates between feedback and no
feedback runs.

The characteristic shape of the PDF in figure~\ref{fig:dpdfz} has been
found in several numerical studies of turbulent ISM. In
particular, it is very similar to the density PDF in the 2D disk
simulations of \citet{wada_norman01}. These authors also found that
density PDF is not sensitive to the stellar energy feedback. The
log-normal high-density tail of the PDF is thought to be a generic
feature of the turbulent (i.e., randomly forced) supersonic flows
\citep[e.g.,][]{vazquezsemadeni94,padoan_etal97,vazquezsemadeni_etal00}.
The effective pressure in such flows is dominated by turbulent
(kinematic) rather than thermal pressure. This may explain the
relatively low sensitivity of the density distribution to the details of
cooling.  The cooling and heating, however, are important in
determining the effective equation of state of the gas and thus should
definitely also play a role. The gas will become increasingly more
compressible as the ratio of the local cooling time to the characteristic
crossing time decreases. Higher compressibility would then allow the gas
to reach higher densities in converging flows, presumably creating the
high-density tail of the distribution and star forming
regions. The high-density tail is thus likely to be the result of the driven
turbulent cascade in a nearly isothermal gas. Indeed, for the
heating/cooling rates adopted in the simulations temperatures of the
gas at densities $\rho_{\rm g}\gtrsim 5\ \rm M_{\odot}pc^{-3}$ are in
the range $\sim 100-1000$~K. The transition to isothermality, however, 
does not explain the shape of the PDF. The same shape was found in a
test simulation in which gas was artificially kept isothermal at $T=300$~K.

\begin{figure}[t]
\centerline{\epsfxsize=3.5truein\epsffile{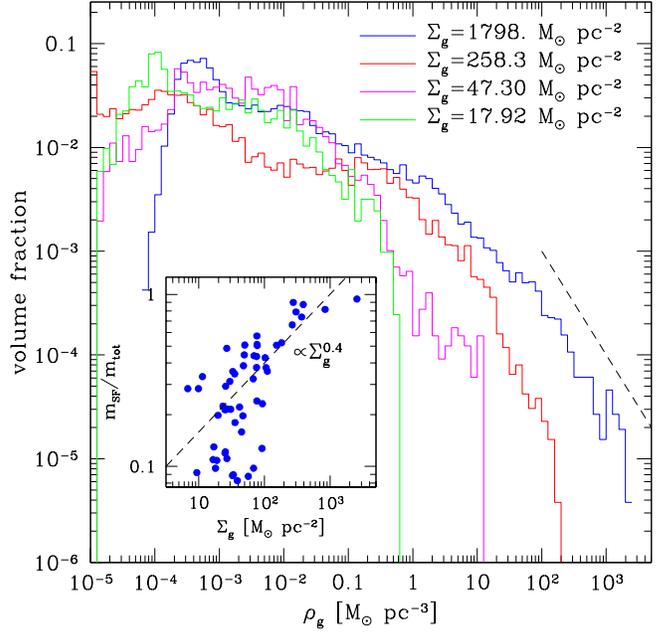} }
\caption{\footnotesize Density PDFs in 3.35 kpc ($l=3$) mesh cells of different surface densities at $z=4$.  The PDFs are normalized to unity.
  The dashed line is $\propto \rho_{\rm g}^{-1}$ dependence.  The
  inset plot shows the fraction of mass at densities $\rho>\rho_{\rm SF}$
  as a function of average surface density of cells.  The dashed line
  here shows scaling $\propto \Sigma_{\rm g}^{0.4}$, expected from the
  Schmidt law (see text for details). }
\label{fig:cdpdf}
\vspace{-0.5cm}
\end{figure}

The Schmidt law on kiloparsec scales implies that the efficiency of the
turbulent build-up of the high-density tail depends on the average
gas density at these scales. Clearly, if the PDF shape were independent of
average density we would have ${\rm SFR}\propto \Sigma_{\rm g}$.
To satisfy the observed non-linear scaling, the fraction of mass in star
forming regions should scale as $m_{\rm SF}/m_{\rm g}\propto \Sigma_{\rm g}^{n-1}$, where $n$ is the slope of the Schmidt law, $m_{\rm g}$ is
the total mass of gas in a volume element used in averaging.

Figure~\ref{fig:cdpdf} shows the density PDFs in $3.35$~kpc ($l=3$)
cells with different densities. Although the low-density part of the
distribution fluctuates, the largest changes are in the high-density
tail which steepens with decreasing gas density. For $\Sigma_{\rm
  g}=18\ \rm M_{\odot}pc^{-3}$ the maximum density is close to the
star formation threshold $\rho_{\rm SF}=1.64\ \rm M_{\odot}pc^{-3}$
adopted in simulations. The inset in the figure~\ref{fig:cdpdf} shows
that the fraction of mass at densities $\rho>\rho_{\rm SF}$ as a
function of average surface density scales as $\propto \Sigma_{\rm
  g}^{0.4}$, as expected from the Schmidt law. Interestingly, this can
also explain the results of \citet{wong_blitz02} who found that ${\rm
  SFR}\propto \Sigma_{\rm H_2}^{n_{\rm mol}}$ with $n_{\rm mol}\approx
1$ while ${\rm SFR}\propto \Sigma_{\rm HI+H_2}^{n}$ with $n\approx
1.4$, where $\Sigma_{\rm H_2}$ is the density of molecular hydrogen.
Most of the gas at pressures $\gtrsim 10^4\ \rm cm^{-3}K$ in their
observations is in molecular form. This corresponds to densities of
$\gtrsim 1-5\ \rm M_{\odot}pc^{-3}$. The molecular gas thus simply
traces the high-density tail of the PDF. Figure~\ref{fig:cdpdf} shows
that constant efficiency of the star formation in the densest regions does
not contradict the Schmidt law for the total gas density on large
scales.

The scatter in $m_{\rm SF}$ for $\Sigma_{\rm g}\lesssim 10-20\ \rm
M_{\odot}pc^{-2}$ increases as the maximum density of the PDF
fluctuates around $\rho_{\rm SF}$ reflecting the stochastic nature of
turbulent flows.  The build-up of the high-density tail can be viewed
as a random walk of gas elements in density with the probability
proportional to the density PDF \citep{elmegreen02}. The bottleneck in
the rate of gas crossing $\rho_{\rm SF}$ is at low densities at
largest scales, which therefore control the mass of gas available for
star formation. Indeed, simulations of \citet{wada_norman01} show that
the density PDF reaches its ``equilibrium'' shape on the dynamical
time corresponding to the average density of the region.

This scenario can naturally explain the tight connection between the
average gas density on kiloparsec scales and star formation on small
scales. The universality of the connection can be attributed to the
relatively low sensitivity of the density PDF to the details of
feedback and precise cooling and heating rates. The efficiency with
which driven supersonic turbulence builds up the high-density tail of
the PDF depends mainly on the average gas density of the region.
Results of this study indicate that the Schmidt law is simply a
manifestation of this dependency.  It is intriguing that the
turbulence may also be responsible for the redistribution of angular
momentum and exponential light profiles of galactic disks
\citep[e.g.,][]{lin_pringle87,silk01,slyz_etal02}.

The modeling of the density dependence and evolution of the PDF
is well within the capabilities of the
current state-of-the-art numerical simulations of the ISM
\citep[e.g.,][]{vazquezsemadeni_etal00,wada_norman01,ostriker_etal01}.
This opens the possibility to construct a sound detailed theoretical
model of the empirical global star formation law in the very near
future.

\vspace{-3mm}
\acknowledgments I would like to thank Anatoly Klypin for many
stimulating discussions on star formation. The simulations and
analyses presented here were performed on the IBM RS/6000 SP system at
the National Energy Research Scientific Computing Center (NERSC) and
on the Origin2000 at the National Center for Supercomputing
Applications (NCSA). This work was supported by the National Science
Foundation under grant No. AST-0206216.

\vspace{-1cm}
\bibliography{../../biblio/data/kravtsov,schmidt}

\end{document}